# Patterns of Learner–AI Interaction and Academic Performance in an Object-Oriented Programming Course

Marina Lepp
*Institute of Computer Science*
*University of Tartu*
Tartu, Estonia
marina.lepp@ut.ee

***Abstract*—This full research paper examines how different forms of learner–AI interaction relate to learning outcomes in object-oriented programming (OOP) courses. Generative artificial intelligence (GenAI) tools are increasingly used by students in programming education, yet evidence on their educational impact remains mixed. In particular, little is known about how students integrate GenAI tools when learning OOP, and how different patterns of use relate to students' learning experiences and outcomes. This study investigates patterns of students' self-directed GenAI use and their relationship with academic performance, perceived difficulty, understanding, and trust. Survey data were collected from 210 undergraduate students enrolled in a first-year OOP course in which the use of GenAI tools was permitted for coursework but prohibited in assessments. Results show that students used GenAI significantly more often for explanation seeking and debugging than for code generation. Cluster analysis identified five distinct learner–AI interaction profiles, including a "smart" high-usage pattern characterized by low reliance on code generation and high use for conceptual support and debugging. While usage patterns were associated with differences in perceived assignment difficulty, self-assessed understanding, trust in AI-generated code, and norm-related attitudes, no significant differences in assessment performance were found across clusters. These findings suggest that self-directed GenAI use alone does not lead to measurable learning gains, underscoring the need for pedagogically guided and process-aware AI support. The study contributes empirical evidence on learner–AI interaction patterns and highlights the importance of pedagogically guided AI use in programming education.**

## I. Introduction

Since the launch of ChatGPT in 2022, the use of Artificial Intelligence (AI) assistants has evolved rapidly, with computer science education being a notable area of impact. Generative AI (GenAI) tools can now generate code, explain programming concepts, identify and correct errors, and provide personalized feedback on programming tasks. These capabilities have the potential to transform the way students learn programming and how educators design instruction. However, students utilize AI assistants in various ways when learning programming, including object-oriented programming (OOP), which presents both opportunities and challenges for learning.

Empirical findings on the educational impact of GenAI tools in programming education remain mixed. Some studies report lower performance among students who use AI assistants [1], [2], whereas others find no significant differences in learning outcomes [3] or even improved performance among users [4], [5]. The effects appear to depend on usage patterns, as engaging with AI for explanations may enhance learning, whereas those who depend on AI-generated solutions may experience hindered problem-solving development [6]. Despite this, evidence suggests that students tend to use AI assistants primarily for coding tasks rather than for seeking explanations or other purposes [3].

Although research on AI-assisted programming education is growing, studies that focus specifically on object-oriented programming remain relatively rare. A recent systematic literature review revealed that only four out of 125 studies have examined the use of AI in OOP courses [7]. This lack of focus is noteworthy, given that OOP is widely considered a threshold concept in computer science education—students often struggle to master it and yet it is crucial for advancing to higher-level programming and design [8]. Understanding how students employ AI tools in this context, and how different interaction patterns affect student performance and skill development in OOP courses, is therefore of particular importance.

## II. Background

The integration of GenAI tools into computer science education has caused widespread interest, particularly in programming courses where these tools offer code generation, debugging assistance, and conceptual explanations. Studies have shown that AI assistants, such as ChatGPT, GitHub Copilot, and Codex, can support students by providing examples, annotations, and alternative problem-solving approaches [9], [10], [11]. Students often perceive AI assistants as helpful resources, especially for coding-related tasks, and some even view them as on-demand tutors [12]. Positive student evaluations are frequently reported for IDE-embedded explainers/tutors and assignment-aligned assistants [11], [13], as well as stepwise solution generators [14]. However, the quality and accuracy of AI-generated content can vary, with issues such as poor code style, minor inaccuracies in explanations, and challenges in interpreting non-textual inputs [15], [16], [17]. While students appreciate the speed and availability of AI tools, their reliance on these assistants varies, often shaped by perceived usefulness, prior experience, and discipline-specific familiarity [10], [18], [19].

Empirical studies on the impact of AI assistants on academic performance present mixed results. While some research reports improved performance, self-efficacy, and computational thinking among students who use AI tools [4], [5], [20], [21], other studies find no significant differences or even negative outcomes [1], [2], [10], [22]. The variation in findings often arises from differences in usage patterns: some students rely mostly on coding-focused behaviors (writing/editing/debugging), whereas others lean toward explanation/ideation [3], [23], [24], [25]. Students who engage with AI for explanations and conceptual support tend to benefit more than those who rely on it for direct solutions [6], [26]. Over-reliance may hinder the development of problem-solving and metacognitive skills, particularly among novice programmers who struggle to critically evaluate AI-generated content [27]. Furthermore, high-achieving students may benefit more from AI, while struggling students risk falling further behind due to over-reliance and reduced cognitive engagement [27]. Usage also scales with task complexity and course level as students in advanced courses or working on complex problems turn to AI more often [28], [29].

Recent research emphasizes the importance of analyzing fine-grained behavioral data to understand how students interact with AI assistants. Patterns such as repeated querying, copy-paste behavior, and passive acceptance of AI suggestions may indicate shallow engagement and hinder the development of problem-solving and metacognitive skills [30], [31]. Concepts like "shepherding" (students rarely produce their own code) and "drifting" (many AI suggestions but little real progress) describe scenarios where students rely heavily on AI without meaningful cognitive engagement, leading to an "illusion of competence" [27], [32]. The implication is clear: instructional support should help students critically examine AI-generated outputs, explain the changes they make, and develop an understanding of the underlying problem-solving processes, rather than simply accepting AI-generated answers [33].

To address these challenges, educators are developing pedagogical frameworks that promote responsible AI use and turn GenAI into a learning instrument rather than a mere solution engine. Recommended approaches include encouraging students to identify and critique AI-generated outputs, utilize AI for personalized introductions to new topics and guided examples, and emphasize the ethical use of AI to prevent violations of academic integrity [31], [33], [34]. To reduce over-dependence on AI and promote reflective learning, researchers suggest process-focused assessment (e.g., requiring students to explain how they evaluated, changed, or rejected AI suggestions), structured reflection on AI outputs ("slow AI" use), and scaffolds that help manage cognitive load by guiding students through problem-solving steps before showing complete code [35], [36], [37], [38], [39]. Furthermore, patterns of AI use are shaped not only by individual behavior but also by task design, rules, and incentives that can encourage or discourage visible help-seeking [40].

From a learning sciences perspective, the educational impact of AI tools may be better understood by considering learners' modes of cognitive engagement rather than usage frequency alone. The ICAP framework distinguishes between passive, active, constructive, and interactive engagement, emphasizing that deeper learning is more likely when learners generate explanations or actively construct understanding [41]. Although ICAP has only recently begun to inform research on GenAI use in programming education [42], it provides a useful lens for interpreting variation in learner–AI interaction patterns. Prior research, therefore, underscores the need for analyses that focus on patterns of learner–AI interaction and their relationship to learning outcomes, particularly in complex domains such as object-oriented programming.

Finally, there is growing recognition that research must move beyond binary comparisons (AI vs. no-AI) to fine-grained, pattern-level analyses that connect what students actually do with AI to what they learn, especially in OOP. By investigating specific patterns of AI assistant usage and their correlation with academic performance in OOP courses, this study aims to contribute to the development of evidence-based strategies that optimize AI integration for meaningful learning outcomes.

## III. Goal

The primary aim of this study is to investigate students' self-reported use of AI assistants and the relationship between different patterns of AI assistant use and academic performance in an Object-Oriented Programming course. The use of GenAI tools during the course was unrestricted but not explicitly encouraged. Throughout the paper, the terms "AI assistants," "GenAI assistants," "AI tools," and "GenAI tools" are used interchangeably to refer to both conversational chatbots (e.g., ChatGPT, Gemini) and integrated coding assistants (e.g., JetBrains AI Assistant, Full Line Code Completion). Academic performance is measured by points earned on major graded assessments—such as programming tests and the final exam—where the use of AI assistants was prohibited, as well as by total course points, which determined the final grade. To address this aim, the study focuses on the following research questions:

1. How do students report engaging with GenAI tools during an OOP course, particularly with respect to code generation, debugging, and explanation seeking?
2. Which distinct patterns of learner–AI interaction can be identified based on students' reported use of GenAI tools?
3. How do different learner–AI interaction patterns relate to students' time investment, perceived task difficulty, self-assessed understanding, and perceptions about AI?
4. What relationships exist between identified learner–AI interaction patterns and students' performance on AI-free course assessments?

By addressing these questions and moving beyond aggregate measures of AI use and performance, this work aims to contribute evidence relevant to artificial intelligence in education by clarifying how learner–AI interaction patterns can inform the design of pedagogically guided, process-aware AI supports in programming education.

## IV. Approach

This study adopts a learner-centered approach to examine how students use GenAI tools in an authentic programming context. Rather than evaluating a specific AI system or intervention, the study focuses on identifying patterns of learner–AI interaction and relating them to learning-related outcomes using self-reported measures. This approach provides insight into how students use AI tools in the absence of explicit instructional guidance.

### *A. Context*

Object-oriented programming (OOP) is one of the most widely used paradigms in modern software development. At the University of Tartu, first-year students are introduced to this paradigm in the course Object-Oriented Programming, which is taught using the Java programming language. The course is a mandatory component of the Computer Science curriculum but also attracts many students from other disciplines, resulting in an annual enrollment of approximately 270–330 students. As the second programming course in the curriculum, it requires prior completion of an introductory Python course [43]. Since no previous experience with Java or object-oriented programming is expected, most students begin the course as complete novices in both the paradigm and the language.

The 16-week course follows a flipped-classroom model [44] and consists of weekly lectures, homework assignments, and practice seminars. Each week, students are expected to watch lecture videos, complete a short quiz, work on topic-related homework, and participate in seminars to deepen their understanding. Most homework tasks include automated tests that provide instant feedback, allowing students to revise and resubmit their solutions. Throughout the course, students complete two major programming tests, two group projects [45], and a final exam conducted in quiz format.

The use of AI assistants is permitted for homework and group assignments. The course materials even encourage students to consult AI tools for explanations and guidance, for example: "If you don't understand the code, you can ask an artificial intelligence (ChatGPT, Copilot, or another tool) to explain the logic," or "You can also ask an artificial intelligence for help in locating the mistake." However, AI assistants are strictly prohibited during programming tests and the final exam. This rule is communicated to students at the start of the course and reiterated before each assessment. During the first lecture, it is also emphasized that students should avoid excessive reliance on AI when completing homework assignments.

### *B. Sample and Data Collection*

The sample comprised 316 students enrolled in the Object-Oriented Programming course in the spring semester of the 2024/2025 academic year, of whom 70.3% were male and 29.7% female. Data on students' use of AI assistants were collected through a survey administered in the 11th week via the course Moodle page. Participation was voluntary, and 210 students responded (66.5% of the cohort), including 134 males (63.8%) and 76 females (36.2%). Although responses were not anonymous to allow linkage with assessment data, analyses were conducted confidentially, and the study followed institutional ethical guidelines with informed consent obtained from all participants.

The survey contained six sections, primarily addressing feedback on lectures, seminars, tests, and homework, with one section dedicated to AI assistants. The questionnaire included three items on students' use of GenAI tools, specifically whether they employed them for code generation, bug identification, or seeking additional explanations. The frequency of AI tool usage in different aspects of the learning process was assessed using a single-choice scale with the options "Did not use," "Once," "A couple of times," "Every other week," "Almost every week," and "Every week." To assess students' perceptions of the use of AI assistants, a set of eight statements was included in the questionnaire: (S1) GenAI tools will negatively impact my future job prospects; (S2) GenAI tools will harm the development of generic or transferable skills such as teamwork, problem-solving, and leadership; (S3) I am concerned that I will become over reliant on GenAI tools; (S4) I trust the code written by GenAI tools more than the code I write; (S5) My instructors can detect code that was written by GenAI tools; (S6) My instructors actively check for unauthorized use of GenAI tools; (S7) If my instructors disallow GenAI tools, it's ok to use them to generate code if I understand the code. It is unethical only if I copy w/o understanding; (S8) If everyone in class is using GenAI tools, but it is against the rules to use them, then I would still use them. Respondents indicated their level of agreement using a 5-point Likert scale ranging from "Strongly disagree" (1) to "Strongly agree" (5). This set of items was adapted from question 13 of the survey conducted by Keuning et al. [29].

In the questionnaire, students were also asked how much time they spent weekly on the course outside of on-site work (a single-choice scale with the options "<4 hours", "4–8 hours", and ">8 hours"). There were also questions about the perceived difficulty of the homework assignments and the first programming test, rated on a 5-point Likert scale ranging from "Too easy" to "Too difficult." In addition, students were asked to rate their understanding of most topics covered in the course by choosing from the following statements: "Blurred understanding of most topics," "I understand the constructs more or less, but too often I can't use them to solve problems," "After some fiddling, I understand the constructs and manage to solve the problems," and "I understand the constructs (relatively) well and can use them to solve problems."

Additionally, performance data were collected for the 210 survey respondents, including scores from Programming Test 1, Programming Test 2, the final exam, and total course points. Each programming test was worth 16 points, with a passing threshold of 12 points for Test 1. The exam carried 33 points, with a minimum passing score of 15. In total, the maximum attainable score for the course was 103 points.

### *C. Data Analysis*

The analysis was performed in Python (pandas, NumPy). Survey responses were analyzed using descriptive statistics. For analysis, the frequency of AI tool usage across different aspects of the learning process was converted into numerical values ranging from 0 ("Did not use") to 5 ("Every week"). Non-parametric Friedman tests were applied to detect overall differences in reported usage, followed by Wilcoxon signed-rank tests with Holm correction for pairwise contrasts.

The Shapiro–Wilk test indicated a significant deviation from normality for all course scores ($W(210) = 0.6–0.87$, $p < 0.001$ in all cases). Given this non-normal distribution and the use of ordinal scales in the survey, Spearman's rank-order correlation was employed to assess the strength of the relationships between course scores and reported usage of AI assistants. Following Akoglu [46], Spearman's correlation coefficients are interpreted as indicating a weak relationship ($r < 0.29$), a moderate relationship ($0.30 \leq r \leq 0.39$), a strong relationship ($0.40 \leq r \leq 0.69$), or a very strong relationship ($r \geq 0.70$).

A cluster analysis was conducted to identify patterns in students' self-reported usage of GenAI tools. Three usage indicators were standardized (z-scores) and used as input for k-means clustering. Students with zero usage across all three indicators were assigned to a separate cluster, while the remaining students were grouped into clusters based on usage frequency. The analysis for non-zero cases tested values of k ranging from 2 to 5. The optimal number of clusters was determined using the silhouette score and inertia. The algorithm used random initialization and converged after 60 iterations. Nonparametric tests (Kruskal–Wallis and Mann–Whitney) were used to compare the identified clusters in terms of the variables used for clustering, Likert-scale responses, and assessment outcomes (two midterm tests, the final exam, and total points). A chi-square test was used to compare distributions between the clusters, such as differences in gender, time spent on the course, and prerequisite course grades.

## V. Results

In total, 27 respondents (12.9%) reported never using AI assistants in this course, while 183 respondents (87.1%) had used them at least once. Among the users, 169 students reported using ChatGPT (Fig. 1). Other popular tools included Full Line Code Completion (47 students), DeepSeek (26 students), JetBrains AI Assistant (26 students), and Gemini (19 students).

Students reported moderate use of GenAI tools overall (Fig. 2). The mean frequency of GenAI tool use for code generation was 2.03, with a median of 2. This value falls slightly below the midpoint of the scale, indicating that although some students used GenAI assistants to generate code, this was not among the most common uses. The standard deviation (SD) of 1.38 reflects moderate variability in how frequently students relied on GenAI for this purpose. In contrast, the mean frequency of using GenAI to seek additional explanations was higher, at 2.75, with a median of 3. This suggests that students were generally more inclined to use AI assistants to deepen their understanding of course material. The corresponding SD of 1.43 indicates slightly greater variability, meaning that while some students frequently turned to AI tools for explanations, others did so rarely. A similar pattern emerged for debugging, with a mean of 2.74 (SD = 1.26, median = 3), suggesting comparable levels of engagement with GenAI assistants for this task as for seeking explanations. Overall, the average frequency of GenAI tool use across all activities was 2.5, with a median of 2.33. This points to a generally moderate level of engagement with AI assistants, while the SD of 1.09 indicates that the degree of GenAI usage did not vary substantially among students. A significant difference in usage frequency across the three purposes was found ($\chi^2(2) = 47.98$, $p < 0.001$). Post-hoc tests showed that code generation was used significantly less often than both debugging and explanation seeking (Holm-adjusted $p < 0.001$ for both). No significant difference was found between the use for debugging and explanation seeking ($p = 0.909$).

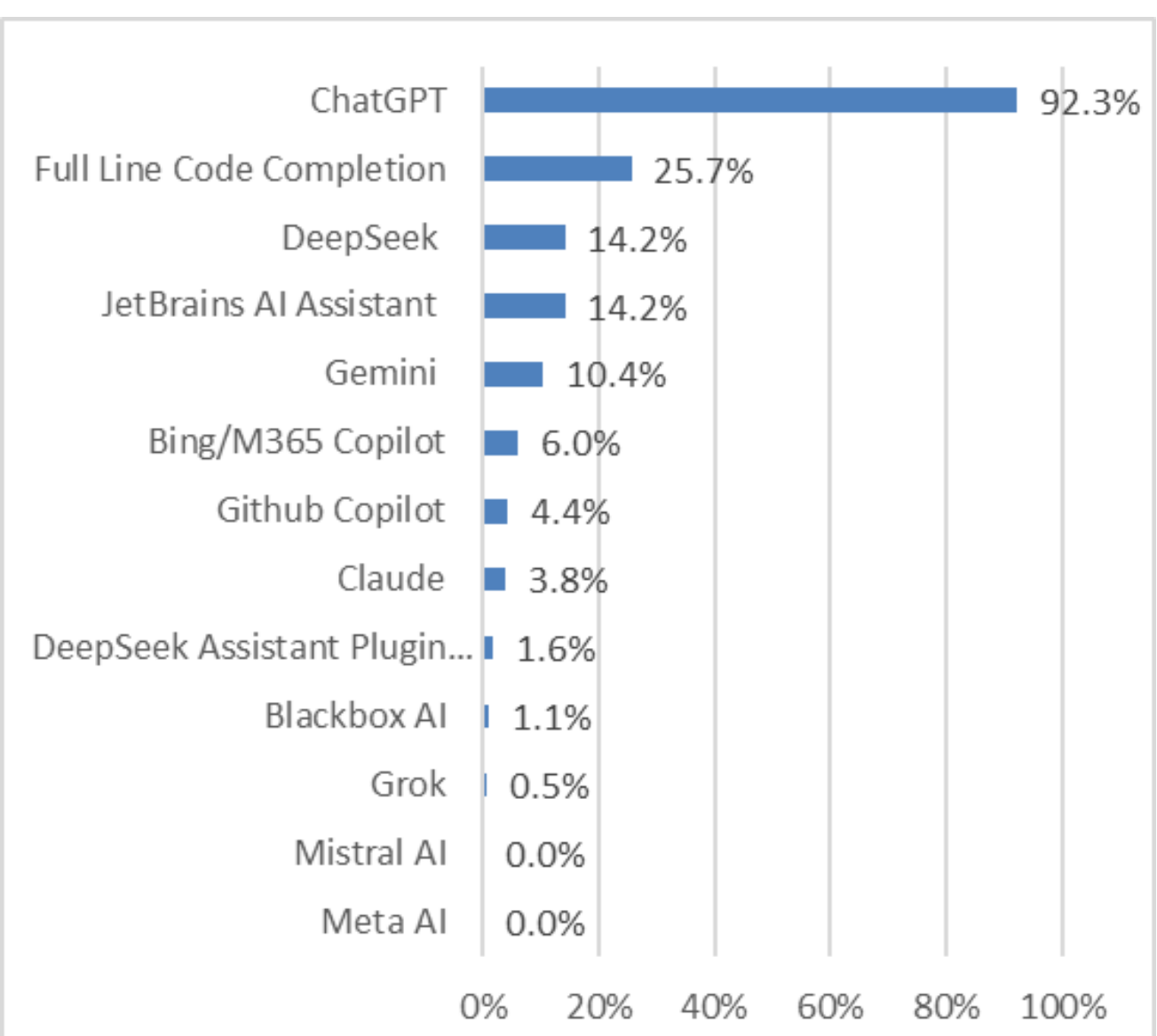

Fig. 1. Reported usage of different AI assistants.

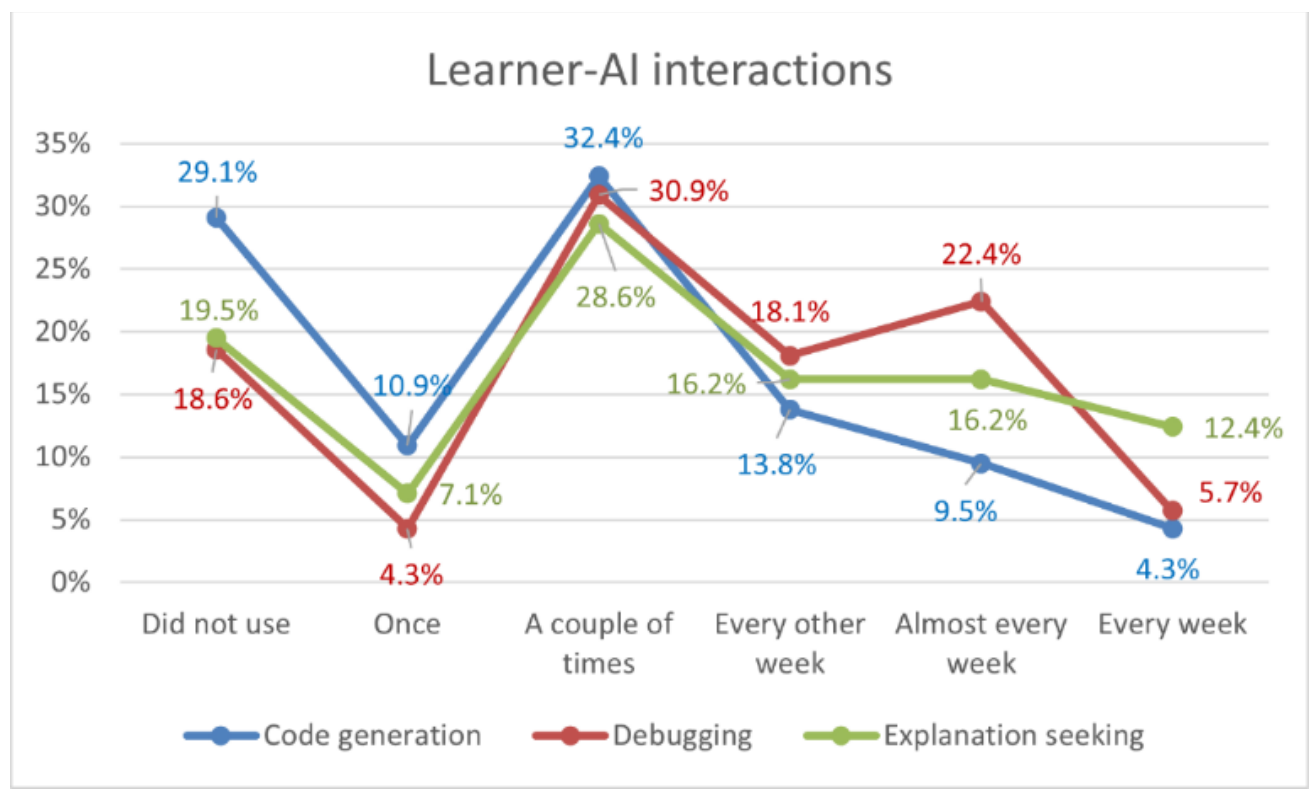

Fig. 2. Reported frequency of the usage of AI assistants for code generation, finding errors, and seeking explanations.

Spearman's correlation analysis revealed only weak negative associations between reported AI use and academic performance (Table I). Although several correlations reached statistical significance, none suggested a strong or consistent relationship between AI use frequency and assessment outcomes.

To further explore usage patterns, a cluster analysis was conducted. After testing solutions with k values ranging from two to five, the four-cluster model was selected as the most interpretable and statistically adequate solution. The model achieved an average silhouette coefficient for the non-zero subset of 0.311, indicating a moderate degree of separation among clusters, and an inertia value of 193.91, reflecting acceptable within-cluster cohesion. Although the silhouette value suggested some overlap between clusters, the resulting profiles were conceptually coherent and practically meaningful. The clusters were characterized as *Zero usage* (n=27), *Low usage* (n=50), *Moderate usage* (n=64), *"Smart" high usage* (low for code generation, high for debugging and explanations; n=32), and *Very high usage* (n=37). Cluster labels are descriptive and refer only to reported usage patterns rather than learner ability or effectiveness. The *"Smart" high*

TABLE I. Correlation Between the Reported Frequency of Different Use Cases and Academic Performance.

| | **Code generation** | | **Debugging** | | **Explanations** | |
|---|---|---|---|---|---|---|
| | ***rho*** | ***p*** | ***rho*** | ***p*** | ***rho*** | ***p*** |
| Test 1 | -0.161* | 0.019 | -0.127 | 0.067 | -0.165* | 0.016 |
| Test 2 | -0.179** | 0.009 | -0.186** | 0.007 | -0.181** | 0.009 |
| Exam | -0.140* | 0.043 | -0.150* | 0.029 | -0.116 | 0.092 |
| Total | -0.169* | 0.014 | -0.147* | 0.034 | -0.138* | 0.046 |

Note: rho = Spearman's rank correlation coefficient; p = significance level. **$p < 0.01$; *$p < 0.05$.

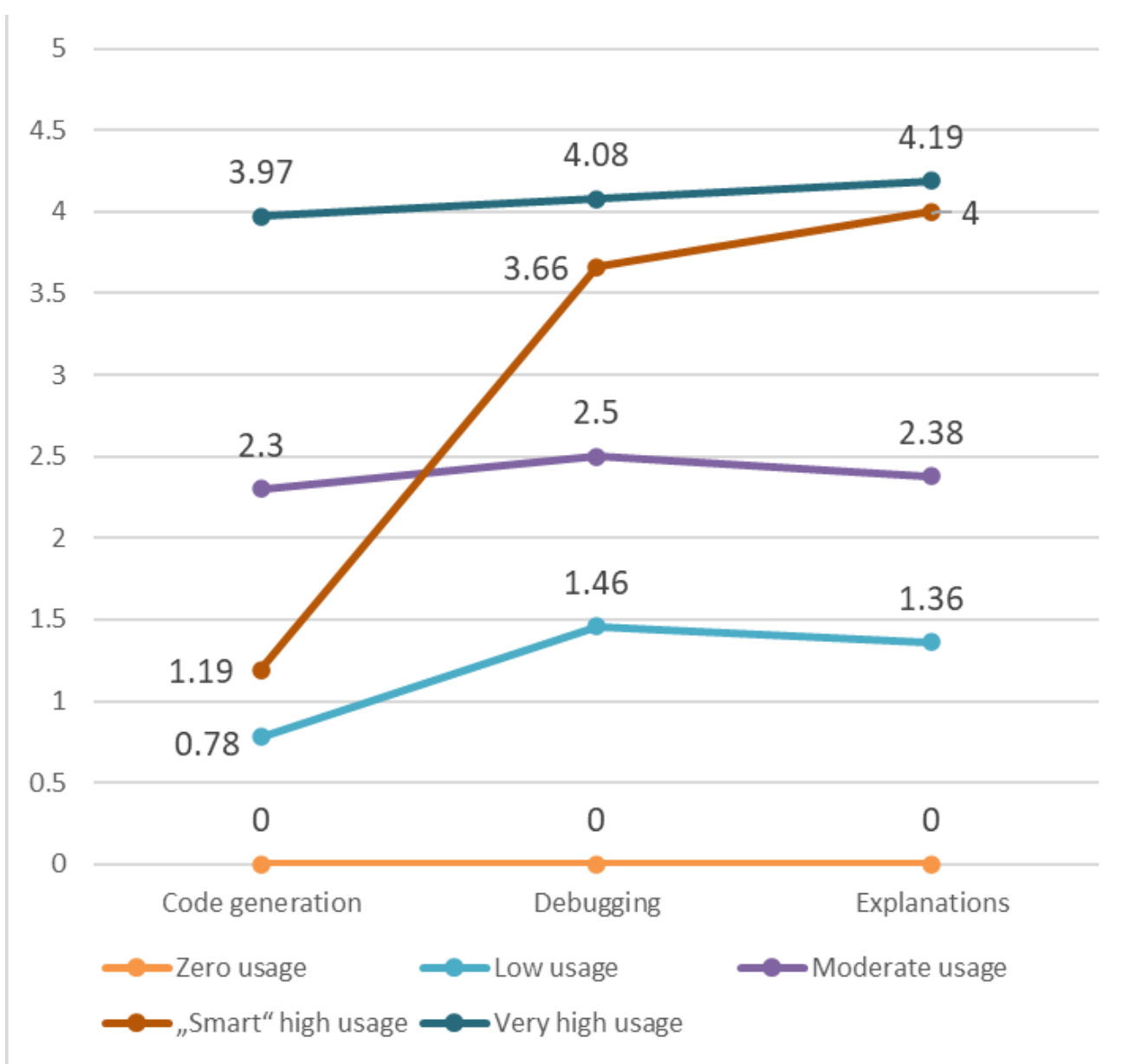

Fig. 3. Mean profiles of AI tool usage across the five user clusters.

*usage* cluster is labeled to reflect terminology used in prior literature, where explanation-focused AI use has been discussed as potentially more supportive of learning than direct solution generation [6], [26].

As shown in Fig. 3, clusters differed clearly in reported usage intensity across all three dimensions. The *Zero usage* cluster reported no engagement with AI tools, while *Low* and *Moderate usage* clusters demonstrated incremental increases in all areas, indicating gradual adoption. The *"Smart" high usage* cluster was distinguished by relatively low code generation (M = 1.19) but high use of debugging (M = 3.66) and explanations (M = 4.00), suggesting a strategic and reflective use of the AI assistants. The *Very high usage* cluster displayed consistently high means across all dimensions (M ≈ 4.0), representing intensive, broad engagement with GenAI tools. These distinct patterns support the interpretability of the five-cluster solution and indicate that the identified groups capture meaningful differences in users' interaction styles with AI assistants.

To further assess the distinctiveness of the cluster solution, Euclidean distances between the cluster centroids were calculated. The smallest distances were observed between the *Low usage* and *Moderate usage* clusters (d = 2.1) and between the *Low usage* and *Zero usage* clusters (d = 2.14). The *Moderate* and *"Smart" high usage* clusters were also relatively close (d = 2.28). In contrast, the largest distance occurred between the *Zero usage* and *Very high usage* clusters (d = 7.07), suggesting strong differentiation between non-users and intensive users. To confirm that clusters differed significantly on the variables used for clustering, Kruskal–Wallis tests were performed for each dimension. Significant differences were found across all three dimensions: code generation, H(4) = 159.01, $p < 0.001$; debugging, H(4) = 157.42, $p < 0.001$; and explanations, H(4) = 154.90, $p < 0.001$. These results indicate that each variable contributed meaningfully to differentiating clusters. The results of pairwise comparisons indicated that the mean ranks of all cluster pairs differed significantly, except for a few comparisons. Specifically, no significant differences were found between *Low usage* and *Zero usage* and between *Low usage* and *"Smart" high usage* in code generation, as well as between *"Smart" high usage* and *High usage* in debugging and explanation seeking.

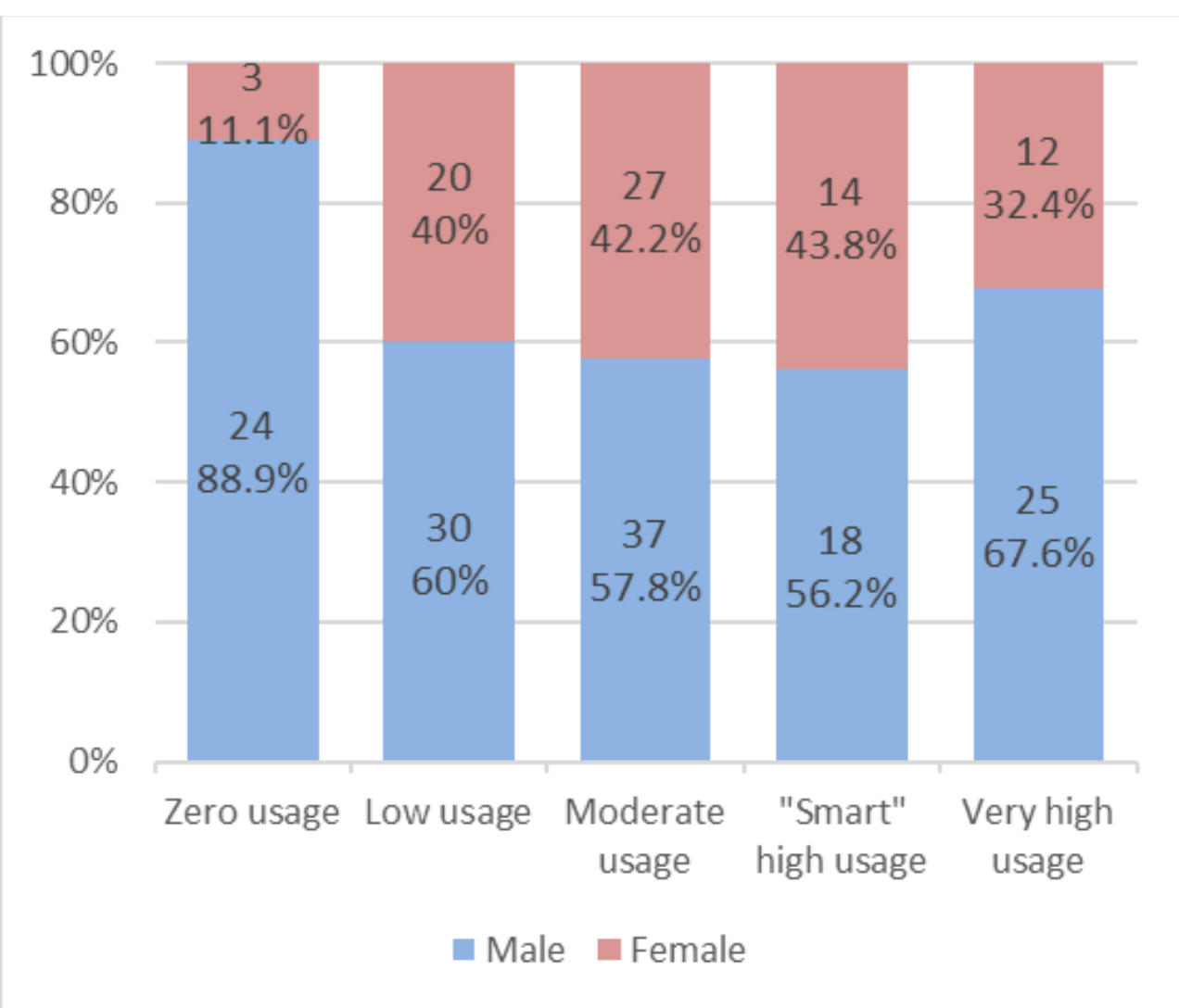

Fig. 4. Gender distribution across the five identified clusters.

To characterize the identified clusters, the proportions of all male and all female participants assigned to each cluster were compared (Fig. 4). An overall association between cluster membership and gender was observed ($\chi^2(4) = 9.68$, p = 0.046). A statistically significant gender difference arose only for the *Zero usage* cluster ($\chi^2(1) = 8.44$, $p < 0.01$): 17.9% of all male participants versus 3.9% of all female participants belonged to this cluster. For the remaining clusters, no significant gender differences were detected (all $p > 0.05$).

In addition, differences in reported weekly time spent on the course among the five clusters were examined (Fig. 5). A chi-square test of independence showed no statistically significant association between time usage and cluster membership ($\chi^2(8) = 8.22$, p = 0.413). To further explore potential trends, a Kruskal–Wallis test was performed using the estimated number of hours as an ordinal variable. The result approached but did not reach statistical significance (H(4) = 8.75, p = 0.068). Although there was a slight tendency for clusters with higher AI usage to report greater time investment, this trend was not statistically significant at the 5% level.

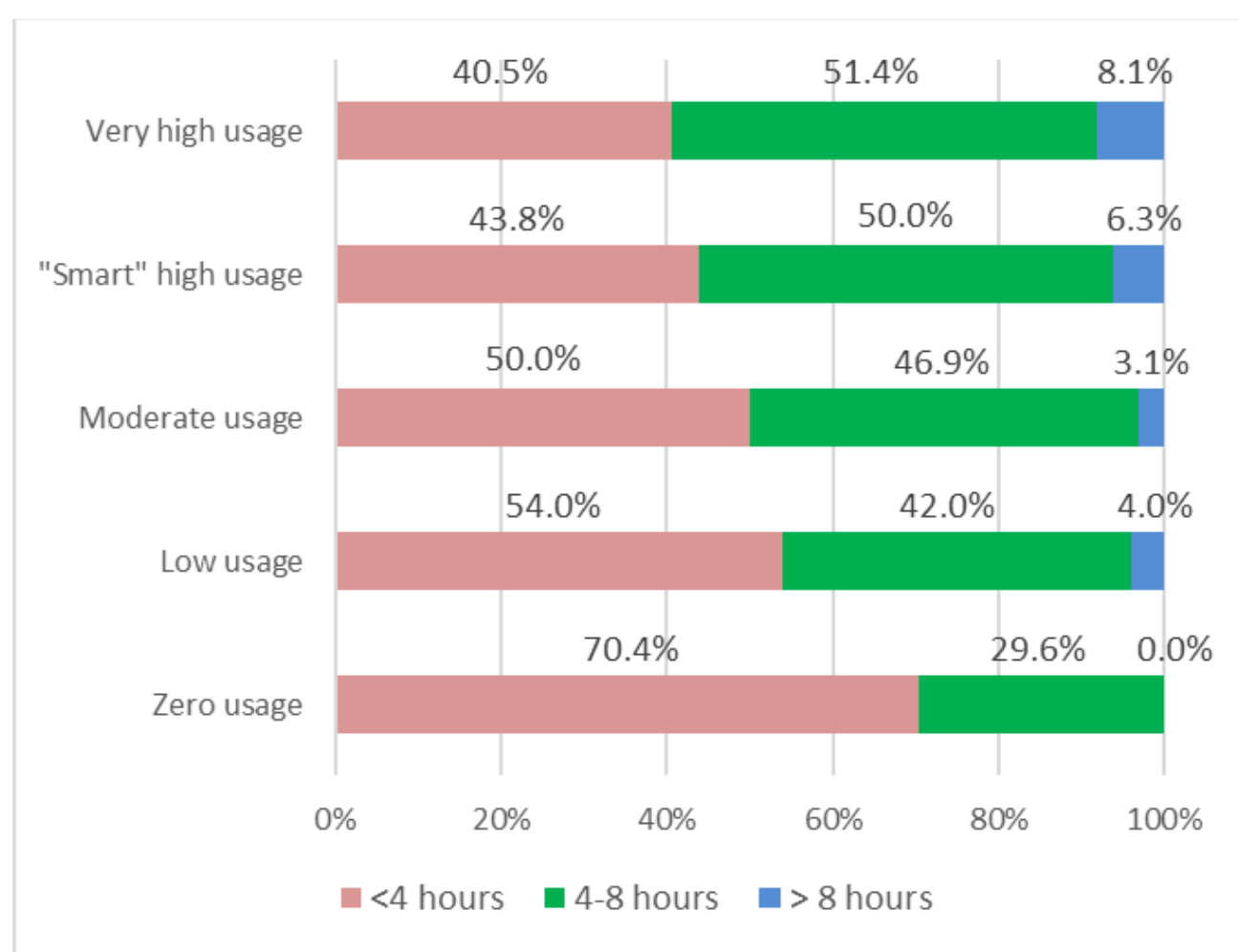

Fig. 5. Weekly time spent on the course across the five identified clusters.

TABLE II. PERCEIVED DIFFICULTY AND UNDERSTANDING IN DIFFERENT CLUSTERS.

| | Zero usage M (SD) | Low usage M (SD) | Moderate usage M (SD) | "Smart" high usage M (SD) | Very high usage M (SD) | H | p |
|---|---|---|---|---|---|---|---|
| Difficulty of homework | 3.15 (0.72) | 3.4 (0.57) | 3.56 (0.64) | 3.56 (0.67) | 3.68 (0.63) | 11.640* | 0.020 |
| Difficulty of test 1 | 2.7 (1.1) | 2.86 (0.53) | 3.06 (0.85) | 3.06 (0.76) | 3.32 (0.71) | 11.965* | 0.018 |
| Unders-tanding | 3.67 (0.62) | 3.48 (0.54) | 3.28 (0.74) | 3.38 (0.71) | 3.19 (0.7) | 10.646* | 0.031 |

Note: H = Kruskal–Wallis test statistic; p = significance level.

*p < 0.05.

Perceived difficulty of homework and the first test also varied across clusters, with higher AI usage associated with slightly higher difficulty ratings (Table II). Specifically, the perceived difficulty of homework in the *Very high usage* cluster differed significantly from that of the *Zero usage* cluster, and the perceived difficulty of the first test in the *Very high usage* cluster differed significantly from the *Zero* and *Low usage* clusters (Mann–Whitney U tests, $p < 0.05$ in all cases). Some differences were also observed in self-assessed understanding: the *Zero usage* cluster rated their understanding significantly higher than the *Very high usage* cluster (Mann–Whitney U test, $p < 0.05$).

In attitudes toward AI tools, significant differences emerged for only two statements (Table III). The fourth statement "I trust the code written by GenAI tools more than the code I write myself" showed the strongest differentiation between clusters. The *Zero usage* cluster reported significantly lower trust than the *Moderate*, *"Smart" high*, and *High usage* clusters (Mann–Whitney U tests, $p < 0.01$ in all cases). Interestingly, the *"Smart" high usage* cluster reported the greatest trust in AI-generated code. Norm-related attitudes regarding peer influence (S8. If everyone in class is using GenAI tools, but it is against the rules to use them, I would still use them) showed smaller but notable differences. The *Zero usage* cluster reported significantly less willingness to use GenAI tools when not allowed compared with the *High usage* cluster (Mann–Whitney U test, $p < 0.05$).

TABLE III. MEAN PERCEPTIONS OF DIFFERENT CLUSTERS.

| | Zero usage M (SD) | Low usage M (SD) | Modera-te usage M (SD) | "Smart" high usage M (SD) | Very high usage M (SD) | H | p |
|---|---|---|---|---|---|---|---|
| S1 | 2.85 (1.23) | 2.9 (1.16) | 2.91 (1.24) | 2.72 (1.25) | 2.81 (1.29) | 0.691 | 0.952 |
| S2 | 3.33 (1.39) | 3.04 (1.19) | 2.89 (1.3) | 2.69 (1.35) | 2.95 (1.31) | 3.566 | 0.468 |
| S3 | 2.96 (1.48) | 3.24 (1.36) | 3.19 (1.07) | 3.06 (1.24) | 3.19 (1.35) | 1.333 | 0.856 |
| S4 | 1.7 (1.14) | 2.4 (1.39) | 2.83 (1.24) | 3.09 (1.25) | 2.62 (1.19) | 20.513*** | 0.000 |
| S5 | 2.74 (0.71) | 3.16 (1.04) | 3.12 (1.19) | 3.06 (1.05) | 3.32 (1.2) | 2.780 | 0.595 |
| S6 | 3.00 (0.88) | 3.14 (1.13) | 3.41 (1.22) | 3.38 (1.13) | 3.16 (1.01) | 4.126 | 0.389 |
| S7 | 2.67 (1.24) | 2.58 (1.33) | 2.64 (1.3) | 2.59 (1.32) | 2.86 (1.27) | 0.890 | 0.926 |
| S8 | 1.89 (1.25) | 2.52 (1.46) | 2.64 (1.36) | 2.47 (1.32) | 2.95 (1.13) | 9.764* | 0.045 |

Note: H = Kruskal–Wallis test statistic; p = significance level.

***p < 0.001; *p < 0.05.

TABLE IV. ACADEMIC PERFORMANCE OF DIFFERENT CLUSTERS.

| | Zero usage M (SD) | Low usage M (SD) | Moderate usage M (SD) | "Smart" high usage M (SD) | Very high usage M (SD) | H | p |
|---|---|---|---|---|---|---|---|
| Test 1 | 14.78 (2.87) | 14.77 (1.40) | 15.10 (1.11) | 14.57 (2.86) | 14.90 (1.07) | 7.647 | 0.105 |
| Test 2 | 13.11 (5.14) | 12.16 (4.87) | 12.53 (4.15) | 12.48 (3.83) | 11.12 (4.34) | 5.281 | 0.260 |
| Exam | 26.59 (7.96) | 24.70 (7.33) | 25.82 (4.94) | 25.25 (5.52) | 25.41 (4.51) | 7.928 | 0.094 |
| Total | 88.50 (18.95) | 85.00 (14.66) | 87.21 (11.46) | 87.67 (12.87) | 84.62 (12.20) | 6.405 | 0.171 |

Note: H = Kruskal–Wallis test statistic; p = significance level.

To examine whether students' prior course performance differed across identified clusters, we compared their prerequisite course grades. No statistically significant differences were found in grade distributions across clusters ($\chi^2(20) = 16.47$, $p = 0.687$) or in average grade ranks ($H(4) = 3.59$, $p = 0.464$). This suggests that cluster membership was not associated with prior academic performance in the prerequisite course. Furthermore, the assessment outcomes of OOP were compared across clusters using the Kruskal–Wallis test (Table IV). No significant differences were found in scores on either of the two midterm programming tests, the final exam, or total course points (all $p > 0.05$).

## VI. DISCUSSION

This study examined students' self-reported use of generative AI assistants in an Object-Oriented Programming course and explored how different usage patterns are related to academic performance. The majority of students reported using GenAI tools at least once, with ChatGPT being the most popular. In response to the first research question, students reported moderate overall engagement with GenAI tools; however, usage was significantly higher for explanation-seeking and debugging tasks than for code generation. This suggests that, in an OOP context, learners primarily employ AI as a sense-making and diagnostic aid, rather than as a code producer. The student can perceive creating code with AI as a risky endeavor because it raises integrity concerns, can produce syntactically correct but semantically shallow solutions, and may not align with grading criteria or coding style and structure expectations [15], [16], [17]. The found patterns align with recommendations to utilize AI for conceptual support and troubleshooting [12], [13], [14] and contrast with the assumption that students mainly rely on AI for solution generation [3].

To answer the second research question, cluster analysis revealed five usage profiles: *Zero usage*, *Low usage*, *Moderate usage*, *"Smart" high usage*, and *Very high usage*, which capture distinct ways learners integrate AI into their workflows. The *"Smart" high usage* cluster stood out for its strategic engagement, characterized by a low reliance on code generation but a high use of AI for debugging and explanation seeking. This pattern aligns with recommendations to question and adapt AI output instead of copying it [33], [35] and supports the literature emphasizing the benefits of explanation-seeking behaviors [6], [26]. Interpreted through the ICAP framework [41], *Zero* and *Low usage* patterns reflect predominantly active engagement without AI mediation, whereas the *"Smart" high usage* pattern aligns more closely with constructive engagement that supports sense-making.

*Very high usage*, in contrast, appears to reflect frequent interaction that remains largely active rather than constructive, where increased activity does not necessarily translate into deeper understanding.

The third research question aimed to determine how usage patterns differ in perceptions and attitudes. Very high users reported greater perceived difficulty of assignments and lower self-assessed understanding, suggesting compensatory help-seeking among students who feel behind or face complex tasks [28], [29]. Interestingly, the *"Smart" high usage* cluster reported the highest trust in AI-generated code, suggesting that reflective use may foster confidence in AI outputs. One possible interpretation is that regularly seeing explanations and identifying errors helps students better judge whether AI-generated code is reliable, leading to more accurate trust in AI while still avoiding full dependence on it [31], [33]. Norm-related attitudes (willingness to use AI when disallowed) varied by cluster as well, indicating that peer influence and perceived classroom norms can shape behavior, potentially overriding formal restrictions. These results underscore the importance of clear guidelines and ethical discussions in preventing misuse and promoting responsible use [34], [40].

Despite these differences in usage patterns and perceptions, neither the frequency of use nor distinct usage profiles were systematically related to academic performance on AI-free assessments, addressing the fourth research question. One possible explanation is that GenAI tools helped students who experienced greater perceived difficulty and lower self-assessed understanding keep pace with their peers, resulting in comparable assessment outcomes across clusters. Interestingly, this interpretation differs from earlier concerns that struggling learners may fall further behind because of over-reliance on AI and reduced cognitive engagement [27]. In addition, because assessments were completed without AI assistance, performance depended on students' independent understanding rather than their ability to effectively use AI during learning activities. Together, these findings suggest that the relationship between GenAI use and academic performance depends not only on how students use AI, but also on how learning activities and assessments are aligned within the course. These results align with emerging research showing that the relationship between GenAI tool use and learning outcomes is complex and context-dependent [1], [3], [4], with benefits that depend not only on usage volume but also on how strategically students engage with such tools [6]. Interestingly, even seemingly smarter uses, such as seeking explanations and debugging, did not improve learning outcomes, in contrast to earlier findings [6]. Importantly, the absence of performance differences does not imply that AI use has no impact on learning; rather, it suggests that self-directed AI use alone may be insufficient to produce measurable gains in grades. These results highlight the need for AI-enhanced learning environments that provide process-aware support, such as prompting reflection, scaffolding problem decomposition, and supporting trust calibration, rather than relying solely on on-demand solution generation. By characterizing learner–AI interaction patterns in an authentic OOP context, this study contributes empirical grounding for the design of pedagogically guided AI support aimed at fostering meaningful engagement rather than superficial or compensatory use.

## VII. Conclusion

This study contributes to the growing literature on AI-assisted learning by providing nuanced insights into how students interact with GenAI tools in an Object-Oriented Programming course and the implications of these behaviors for learning outcomes. While most students reported using AI assistants, usage patterns varied significantly, with a strategic approach focused on debugging and explanations emerging as a potentially beneficial use. Importantly, no significant differences in academic performance were found across usage clusters, suggesting that self-directed use of GenAI in programming courses neither enhances nor hinders assessment results. In addition, students who used AI tools more often reported higher trust, but also greater perceived difficulty of assignments and lower understanding, indicating a complex relationship between AI use, confidence, and learning.

These findings have several implications for engineering and computing education in the era of AI. Simply allowing access to GenAI tools is unlikely to improve learning outcomes without instructional support that promotes responsible, thoughtful, and ethical use of AI. Educators should focus on how students use these tools by encouraging explanation-seeking, critical evaluation of AI-generated outputs, and reflection on problem-solving processes. Incorporating activities such as requiring students to justify AI-assisted solutions or compare AI-generated and self-developed code may help promote deeper engagement. More broadly, the results suggest that effective integration of GenAI tools requires pedagogical strategies that guide students toward meaningful use, rather than relying on unguided or purely task-oriented interactions.

**Limitations.** While this study provides valuable insights into students' use of GenAI tools in an OOP course, several limitations should be acknowledged. First, the data relied on self-reported measures, which may not fully capture actual behavior and may be subject to recall bias or social desirability effects, especially regarding rule compliance and perceived understanding. Second, the study did not track actual AI interactions or behavioral data, limiting the ability to assess the depth and quality of engagement with AI tools. Because cluster formation was based entirely on these self-reported measures, the identified clusters should be interpreted as perceived usage profiles rather than objective representations of students' actual interactions with AI tools. Third, the course context, where AI use was unrestricted but not explicitly encouraged, may have influenced usage patterns differently than in courses with clear AI integration policies. Fourth, the sample was drawn from a single institution and course, which may limit the generalizability of the findings to other educational settings or programming paradigms. Fifth, while cluster analysis revealed meaningful usage profiles, the moderate silhouette coefficient suggests some overlap between clusters, indicating that the identified groupings capture broad tendencies rather than sharply separable categories, and student behaviors may not fit neatly into discrete categories. In addition, because engagement depth was not directly measured and some usage clusters were relatively small, the study may have been underpowered to detect small performance differences; therefore, observed null effects should be interpreted cautiously. Finally, because the study was observational rather than experimental, the findings should be interpreted as descriptive associations rather than evidence of causal effects of AI use on learning outcomes.

**Future work.** Future work should prioritize process-level measures of learner–AI interaction (e.g., prompt revisions, explanation requests, and response edits) to better capture differences between active and constructive engagement. Building on the findings of this study, future research should incorporate behavioral data such as actual AI interaction logs, query types, and response modifications to better understand the depth and quality of student engagement with GenAI tools. Longitudinal studies could investigate how AI usage evolves over time and its impact on learning, especially as students progress from novice to advanced programming tasks. Additionally, experimental designs could test the effectiveness of pedagogical interventions, such as scaffolding, reflection prompts, and process-focused assessments, in promoting the meaningful use of AI. Research should also examine how individual differences (e.g., self-efficacy, cognitive styles) interact with AI usage patterns to influence learning outcomes. As gender differences were minimal, except in the *Zero* usage cluster, which had a higher proportion of male students, this may reflect differing attitudes toward AI adoption, but further research is needed to explore the underlying causes. Finally, expanding the scope to include diverse institutional contexts, course formats, and programming paradigms would enhance the generalizability of the findings and support the development of inclusive, evidence-based strategies for integrating AI in computing education.

## Acknowledgment

This work was supported by the Estonian Research Council grant "Developing human-centric digital solutions" (TEM-TA120).